# The Evidence of Cathodic Micro-discharges during Plasma Electrolytic Oxidation Process

A.Nominé [1,2 a], J.Martin[1], C.Noël[1], G.Henrion[1], T.Belmonte[1], I.V Bardin[2], V.L Kovalev[2], A.G Rakoch[2]

[1] *Institut Jean Lamour – UMR 7198 CNRS – Université de Lorraine – Parc de Saurupt – 54011 Nancy, France*

[2] *National Institute of Science and Technology "MISiS", 4, Leninskij Prospekt, Moscow 119049, Russia*

**Abstract**

Plasma electrolytic oxidation (PEO) processing of EV 31 magnesium alloy has been carried out in fluoride containing electrolyte under bipolar pulse current regime. Unusual PEO cathodic micro-discharges have been observed and investigated. It is shown that the cathodic micro-discharges exhibit a collective intermittent behavior which is discussed in terms of charge accumulations at the layer/electrolyte and layer/metal interfaces. Optical emission spectroscopy is used to determine the electron density (typ. $10^{15}$ $cm^{-3}$) and the electron temperature (typ. 7500 K) while the role of $F^-$ anions on the appearance of cathodic micro-discharges is pointed out.

Plasma Electrolytic Oxidation (PEO) is a promising plasma-assisted surface treatment of light metallic alloys (*e.g.* Al, Mg, Ti). Although the PEO process makes it possible to grow oxide coatings with interesting corrosion and wear resistant properties, the physical mechanisms of coating growth are not yet completely understood.

Typically, the process consists in applying a high voltage difference between a metallic piece and a counter-electrode which are both immersed in an electrolyte bath. Compare to anodizing, the main differences concern the electrolyte composition and the current and voltage ranges which are at least one order of magnitude higher in PEO[1]. These significant differences in current and voltage imply the dielectric breakdown and consequently the appearance of micro-discharges on the surface of the sample under processing. Those micro-discharges are recognized as being the main contributors to the formation of a dielectric porous crystalline oxide coating.[2]

Nevertheless, the breakdown mechanism that governs the appearance of those micro-discharges is still under investigation. Hussein *et al.*[3] proposed a mechanism with three different plasma formation processes based on differences in plasma chemical composition. The results of Jovović *et al.*[4,5] concerning physical properties of the plasma seem to corroborate this mechanism, and also point out the importance of the substrate material in the plasma composition.[6]

Compared with DC conducted PEO process, using a bipolar pulsed DC or AC current supply gives supplementary control latitude through the current waveform parameters. The effect of these parameter on the micro-discharges behavior has been investigated in several previous works.[2,3,7,8] One of the main results of these studies is the absence of micro-discharge during the cathodic current half-period.[9-11] Even if the cathodic half-period has an obvious effect on the efficiency of PEO as well as on the coating growth and composition, the micro-plasmas appear only in anodic half-period. Sah *et al.*[8] have observed the cathodic breakdown of an oxide layer but at very high current density (10 $kA.dm^{-2}$), and after several steps of sample preparation. Several models of micro-discharges appearance in AC current have already been proposed.[1,2,8,12,13]

Though cathodic micro-discharges have never been observed within usual process conditions, the present study aims at defining suitable conditions to promote cathodic micro-discharges and at studying the main characteristics of these micro-plasmas.

[a] alexandre.nomine@univ-lorraine.fr

The process was conducted on commercial EV31 magnesium and Al 2024 aluminum alloys substrates in electrolytes composed exclusively of $NH_4F$ or KOH diluted in deionized water with a concentration of 0.27 mol.$L^{-1}$ and 0.036 mol.$L^{-1}$ respectively. Prior to process, samples were ground down to 1200 SiC paper and successively rinsed in distilled water and ethanol. The coatings were grown using a bipolar pulse current generator (Ceratronic®), which was set to deliver a symmetrical 100Hz quasi-square current waveform. Both cathodic and anodic amplitude were set at 8 A which corresponds to a current density of 20 A.$dm^{-2}$ supplied to the sample.
The global light emitted by the micro-discharges was directly collected by a Hamamatsu R928 photomultiplier (PM) whose output signal was amplified through a 300MHz bandwidth current amplifier (Standford Research Systems SR445).
As a complement, the number of micro-discharges was determined from the analysis of images recorded using a Photron Fastcam SA1.1 ultra-high speed video camera (UHSC) whose acquisition rate was set at 125 kHz. Using the TRACE software,[14] the analysis of the UHSC video images allowed us to estimate the number of micro-discharges over the sample surface per current period during the process.
The chemical composition and the electron parameters of micro-discharges were investigated by means of optical emission spectroscopy using a 550 mm focal length monochromator (Jobin-Yvon TRIAX 550) equipped an ICCD detector. For chemical investigation a 100 gr/mm grating was used, while the recording of the Hα line profile was achieved using a 1200 gr/mm grating.
Spectra acquisition as well as fast video recording were synchronized with the current pulses and regularly acquired over an entire period of the current pulses. Current, voltage and PM output signals were collected and recorded using a 1 GHz bandwidth oscilloscope (Agilent 54832B).

Measurement of the whole light emitted by the micro-discharges (with no spectral resolution) clearly shows a dependence of the micro-discharges appearance on the substrate nature and the electrolyte composition. Within KOH electrolyte, micro-discharges appear on aluminum during the anodic half-period while no micro-discharge was observed on Mg substrate (Fig. 1(a)). Using $NH_4F$ electrolyte, micro-discharges were detected on the surface of Mg substrate during the cathodic half-period while no light emission was detected on the Al substrate surface (Fig. 1(b)).
The light emission signals show that cathodic micro-discharges appear suddenly as the current switches from the positive to the negative values. Then the emitted light intensity decreases gradually with a pseudo-periodic oscillating behavior. The analysis of UHSC films shows a collective behavior of micro-discharges that switch off and on with a pseudo-frequency in the range 8-12 kHz. It is worth noting that the cathodic micro-discharges occur only during 34-42% of the cathodic half-period (Fig. 1(b)) while anodic micro-discharges are observed during the whole positive half-period (Fig. 1(a)). Spectrally resolved measurements point out that the behavior of cathodic micro-discharges does not depend on the observed species (Fig. 1(c)). Indeed elements from either the metallic substrate (Mg, $Mg^+$) or the electrolyte (H) or the growing layer (MgF) exhibit the same variation.

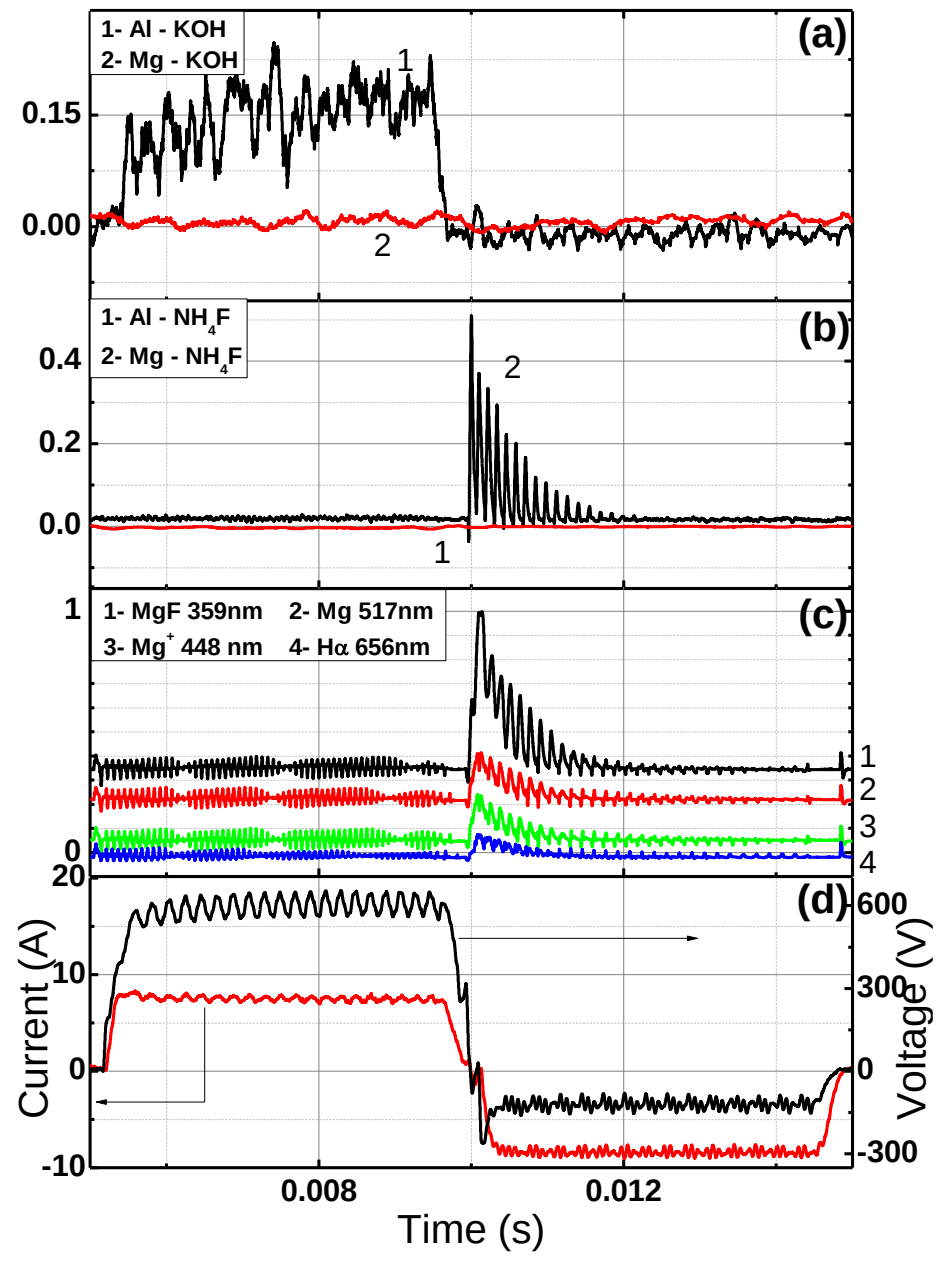

*FIG. 1 Time-resolved variation of electrical parameters and light emission over one period of the current.*
*(a) light emission integrated over the whole spectrum range for Al and Mg substrates in KOH electrolyte.*
*(b) same as (a) in $NH_4F$ electrolyte.*
*(c) line intensity at various wavelengths.*
*(d) applied current waveform and resulting voltage.*
*Vertical axis for Figs. (a) to (c) is emission intensity in arbitrary unit.*

The variation of spectral line intensity with the process time (Fig. 2(a)) shows that all species have similar trend as the process progresses.

The variation of voltage amplitude over the process time is plotted in Figs. 2(b)-(c). From the analysis of Figs. 1 and 2, it can be deduce that there exists a threshold value of the voltage below which no micro-discharges is ignited. Some authors estimated this threshold to about 320-350 V for aluminum substrate in alkaline electrolyte[15,16] which could explain that no breakdown occurs on Mg in KOH. In the case of $NH_4F$, which is a strongly electronegative electrolyte, electrochemical processes are different. This results in discharge breakdown during the anodic half-period of the current with a threshold negative voltage greater than ≈70 V (Fig. 2(b)). Therefore no breakdown occurs on Al in $NH_4F$ while sparks are clearly observed on Mg in the same electrolyte. This difference could be due to a difference in electrolyte conductivity and charge carrier nature.

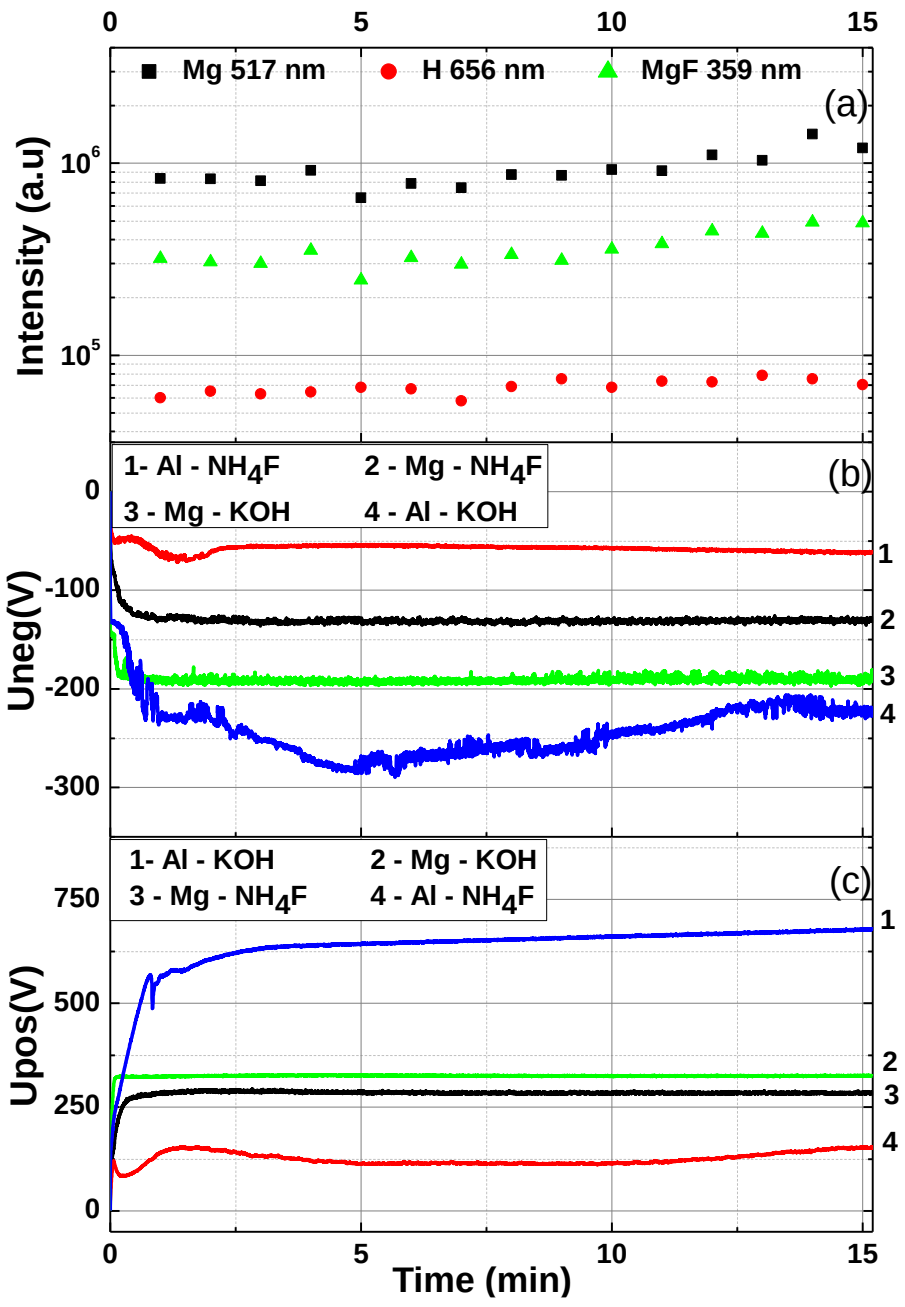

*FIG. 2. Evolution of (a) spectral line intensity within $NH_4F$ electrolyte, (b) negative and (c) positive voltage amplitude during the PEO process on different substrates and electrolytes.*

The variation of discharge parameters with the process time is reported in Fig. 3. Each data point in Fig. 3(a) corresponds to the average number of micro-discharges calculated from images recorded over 8 successive periods of the current (80 ms). Fig. 3(a) clearly shows that the number of micro-discharge per period of the current decreases during the process from 600 at the beginning to 200 after 40 min processing.

A deeper analysis of chronograms (as Fig. 1(d)) recorded every minute over the 40 minutes of the process allowed us to plot the evolution of the current during the presence of micro-discharges (Fig. 3(b)) and their global light emission (Fig. 3(c)). The calculation method consists firstly in sequencing the light-emission signal obtained by PM in N emission peaks separated by breaks. Each light emission peak is then integrated over $[i^- ; i^+]$ which corresponds to the emission time interval of peak i. Finally, all integrated values over one period of the current are summed to give the total emission (ε) per current period (Fig. 3(c)):

$$\varepsilon = \sum_{i=1}^{n} \int_{i-}^{i+} S dt$$

where S represents the PM signal, t is the processing time and N the number of emission peaks in one period, which is equal to 16±2. Simultaneously, for each emission peak the current <I> averaged over the same $[i^- ; i^+]$ interval has been measured and is given in Fig. 3(b):

$$< I >= \frac{\sum_{i=1}^{n} \int_{i-}^{i+} Idt}{\sum_{i=1}^{n} \int_{i-}^{i+} tdt}$$

where *I* represents the current measured by oscilloscope as in Fig. 1(d). Despite the fact that the number of micro-discharges per period decreases significantly (Fig. 3(a)), the integrated light emission increases by about 50%. Since the average current during the presence of micro-discharges is constant, we can suppose that the current per micro-discharge increases and therefore changes the physical properties of micro-plasmas, especially the electron temperature ($T_e$) and electron density ($N_e$).

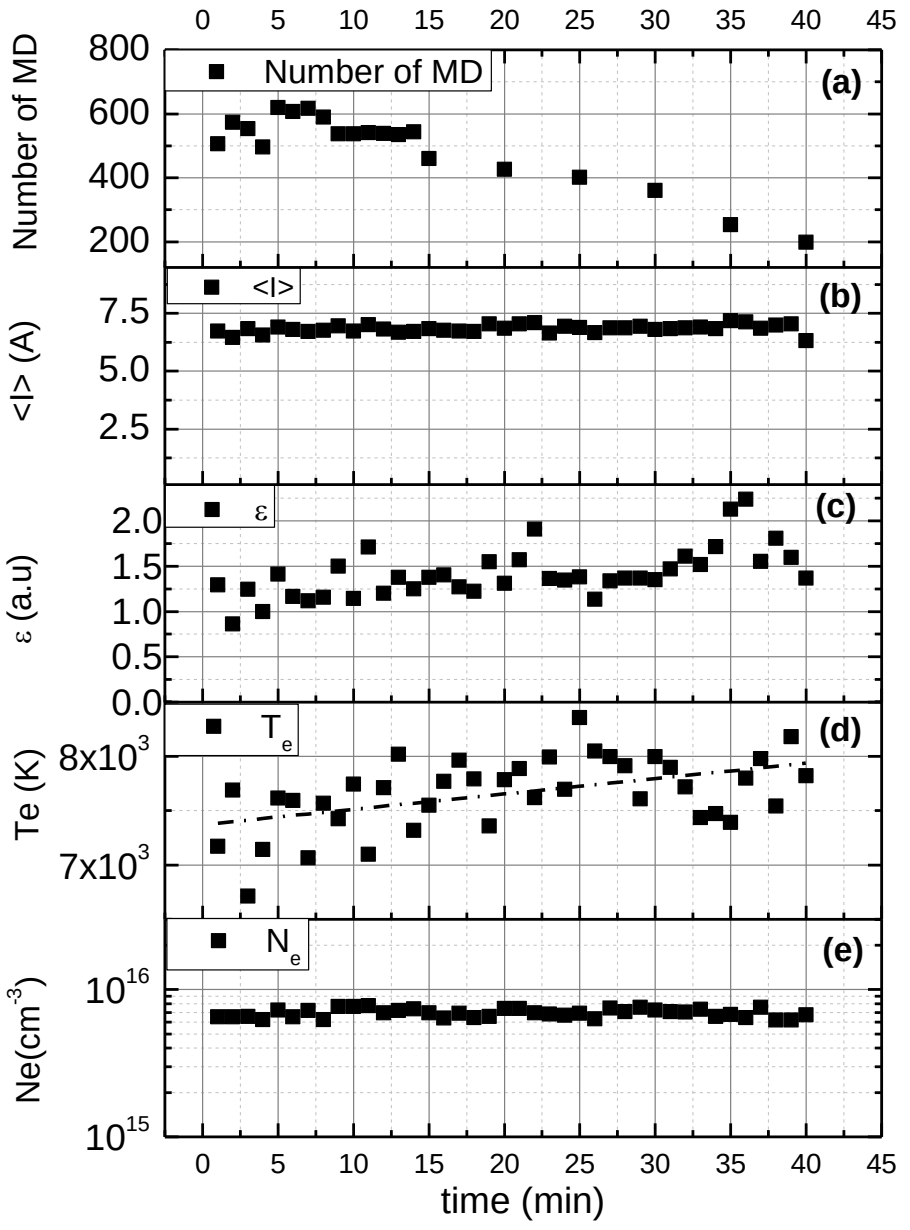

*FIG. 3. Evolution of micro-discharge parameters with the process time. (a) number of micro-discharge per current period; (b) time-averaged current during the presence of micro-discharges; (c) wavelength-integrated light emission intensity; (d) electron temperature; (e) electron density*

Low resolution emission spectra of micro-discharges (Fig. 4), reveals the presence of substrate atoms and ions, products of substrate oxidation (MgF, MgO) and hydrogen. It is worth noting that alloying element (Zn, Zr, Nd, Gd) are not observed in Fig. 4. This is likely due to their low amount in the base material (0.16, 0.3, 0.44, 0.24 at% respectively). Moreover, the high melting temperature of Zr, Nd and Gd does not allow precipitates of these elements to evaporate during PEO.[17] Zn was also not detected since the most intense line (213.8 nm) which corresponds to a resonant transition is out of the spectral range allowed for our equipment. Although all the experiment is regularly cleaned and rinsed, sodium traces from previous electrolytes still remain, and sodium is easily excited to the $^2S_{1/2}$ levels by

**Отформатировано:** Шрифт: (по умолчанию) Times New Roman, 12 пт

**Отформатировано:** Шрифт: (по умолчанию) Times New Roman, 12 пт

electron collision (2.1 eV) and relaxes to the ground level (resonant transition) with a high transition probability of A=6.15 $10^7$ $s^{-1}$,[18] which explains the presence of the 589 nm Na line.

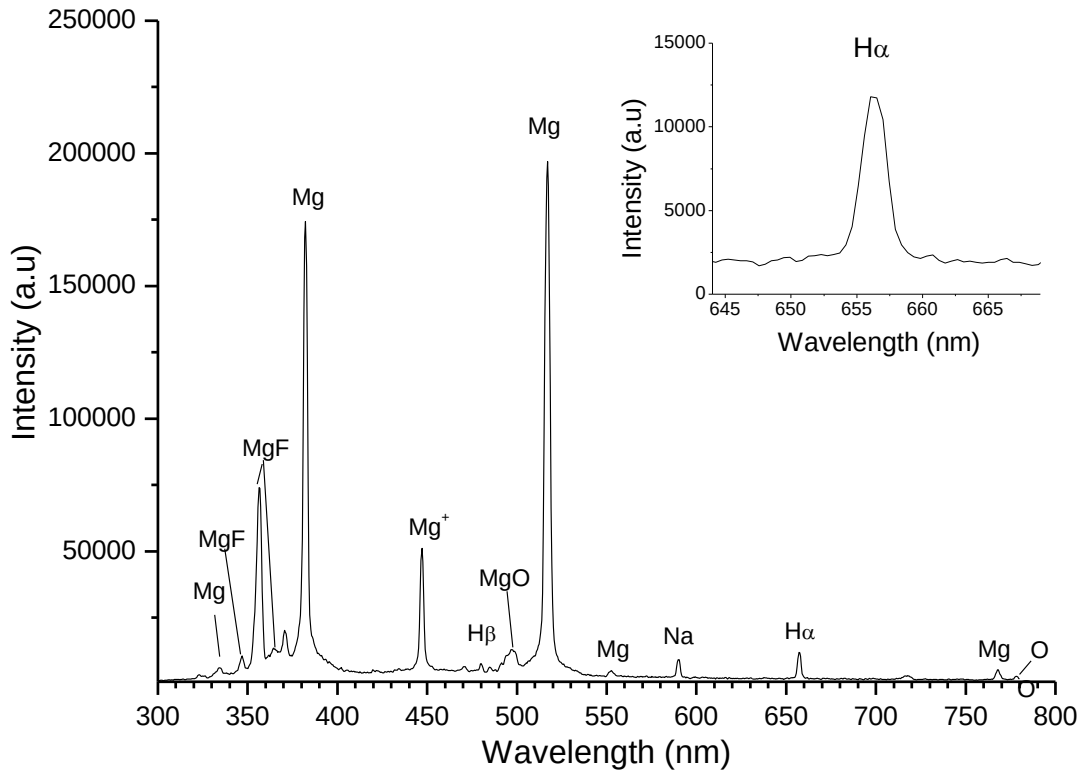

*FIG. 4. Low resolution optical emission spectrum at 40 min. of treatment. Inset represents a high resolution $H_\alpha$ line profile.*

Boltzmann plot of four Mg I lines (330 nm, 383nm, 517nm, 552 nm) with quite wide range of energy levels (1.5 eV) and good dispersion was used to determine the electron temperature $T_e$. Statistical weight $g_i$ and transition probability $A_i$ are taken from Wiese *et al.*[18]. $N_e$ was estimated from the profile of the $H_\alpha$ line recorded using the 1200 gr/mm grating (inset in Fig. 4). The FWHM of the instrumental profile is 0.08 nm while the maximum FWHM due to Doppler broadening of the $H_\alpha$ line is 0.04 nm corresponding to an H-atom temperature of 7500 K. Therefore, the line profile has been fitted using a Voigt function with a fixed Gaussian FWHM of 0.09 nm. The Lorentzian component of the line profile consists of both Stark and Van der Waals broadening. According to Yubero et al.[19] the Van der walls FWHM can be estimated by:

$$\Delta\lambda_w(H_\alpha) = 5.736T^{-0.7}$$

where T is the gas temperature. The largest $\Delta\lambda_w(H_\alpha)$ may therefore be estimated at 0.05 nm assuming T > 900 K. This temperature range is confirmed in the case of PEO processing of aluminum alloys which exhibit the $\alpha$-$Al_2O_3$ phase which can only be achieved at high temperature (> 1000 K).[20] Hence the electron density has then been deduced from the Stark component of the Lorentzian contribution using the following the formula given by Descoeudres [21] based on Gigosos' theory: [22]

$$N_e = 8.8308\times10^{16}\times(\Delta\lambda_{wL})^{1.6005}$$

where $\Delta\lambda_{wL}$ is the Stark width of the line. If no significant evolution of the electron density was detected (Fig. 3(e)), a slight increase in the electron temperature can be observed (Fig. 3(d)).

First of all, it is important to point out the different behavior of micro-discharges in fluoride ammonium electrolyte compare to conventional electrolytes (e.g KOH electrolyte). Moreover, the optical investigation of micro-discharges showed unusual results as compared with the use of conventional electrolytes. Therefore this suggests that the breakdown mechanisms in the present case are different than those usually described for PEO.

The appearance of cathodic micro-discharges is at least as surprising as the absence of the anodic ones. However the oxide coating growth still occurs during the anodic half-period of the current due to oxidation reactions at anode:

$$Mg^{2+} + O^{2-} \rightarrow MgO$$

$$Mg^{2+} + 2F^{-} \rightarrow MgF_2$$

Since the formed magnesium oxide layer is a dielectric ($\varepsilon_r \leq 10$),[23,24] electrical conduction is rapidly stopped as the coating grows and consequently the oxidation reactions stop too. As the current polarity changes from positive to negative values (cathodic half-period) negative charges accumulate at the metal/insulator interface while cations (likely $NH_4^+$) accumulate at the layer/electrolyte interface. The layer breakdown depends on the electric field across the ceramic coating and thus on the accumulated charges on each side of the layer. Consequently the time behavior of the micro-discharges follows a succession of charge accumulation steps and breakdown events. This explains the quasi-periodic emission of light (Fig. 1(a)). Indeed, lightning phases correspond to the breakdown event while there is no light emission during the charge accumulation.

Meanwhile the bursts of breakdown result in opening pores across the dielectric layer, which fill with electrolyte. This restores the electrical connection between the electrolyte and the metallic substrate. Thus as the cathodic time goes on (over one cathodic half-period), such electrical connections get progressively more numerous which results in a decrease in the breakdown events, and hence in the light emission until complete extinction at about 1.7 -2.1 ms after the polarity inversion. During the following anodic half-period of the current, the oxidation mechanisms take place again and the process repeats like this until the end of the PEO treatment.

It is worth noting that the presence of MgO and $MgF_2$ has been confirmed by XRD analyses of the coating. Moreover, optical emission spectroscopy of the micro-discharges showed the presence of MgO and MgF molecules issued from the ceramic layer, which confirms our hypothesis of the partial destruction of the layer by the breakdown events. However, the electron density value measured from the Stark broadening of the $H_\alpha$ Balmer line (Fig. 3(e)) is not consistent with an electrical breakdown in the solid for which the electron density should be at least some $10^{19}$ $cm^{-3}$. It should be considered that the breakdown events take place in the electrolyte-impregnated layer which is more consistent with the electron density value of 6-9x$10^{15}$ $cm^{-3}$.

Finally the above discussion could apply whatever the electrolyte composition and the material if one consider that the negative charges involved in the accumulation process are electrons. However, in the present conditions it is likely that the main negative charges that play a role in the mechanisms described above are fluorine negative ions $F^-$. Therefore phenomena within these particular conditions of electrolyte are driven by negative ions instead of electrons which results in the difference observed. Moreover, since $F^-$ ions consume electrons, this would also explain the low electron density value reported in Fig. 3(e).

The following conclusions can be drawn from this work.

a) Micro-discharges during cathodic half-period have been observed on Mg-alloy in an $NH_4F$ containing electrolyte.
b) A collective oscillating behavior of cathodic micro-discharges appearance has been observed.

c) From hygrogen $H_\alpha$ line, the electron density has been determined ($N_e \sim 6\text{-}9 \times 10^{15}$ cm$^{-3}$). Such a value suggests that the breakdown events do not occur in the solid but in the electrolyte-impregnated ceramic layer.
d) From Boltzmann plot using magnesium lines, electron temperature value of ~ 7000 - 8000 K was determined.
e) The unusual observation and behavior of PEO cathodic micro-discharges has been discussed in terms of successive steps of charge accumulations and breakdown events. It was suggested that fluorine negative ions are the main negative charges involved in the observed phenomena.

**Acknowledgement**

We greatly acknowledge the Conseil Régional de Lorraine for granting A. Nominé's PhD work under decision 11CP-769. We are also grateful to Liebherr SA, Eurocopter, Turbomeca and Agusta Westland companies for their contribution to this work in the frame of the Coproclam project (grant 270589) supported by the European Union within the Clean Sky JTI program.